\documentclass[12pt]{iopart}
\usepackage{epsf}
\begin{document}

\review[Interacting electrons in nanotubes]
{Spin-orbit coupling and electron spin resonance for interacting
electrons in carbon nanotubes} 

\author{A De Martino\dag\ , R Egger\dag\ 
\footnote[3]{To whom correspondence should be addressed 
(egger@thphy.uni-duesseldorf.de)},
F Murphy-Armando\ddag\ and K Hallberg\ddag\
}

\address{\dag\ Institut f{\"u}r Theoretische Physik, 
Heinrich-Heine-Universit{\"a}t, D-40225 D{\"u}sseldorf, Germany}

\address{\ddag\   Instituto Balseiro, Centro At\'omico Bariloche, Comisi\'on 
Nacional de Energ\'{\i}a At\'omica, 8400 S.C. de Bariloche, Argentina}

\begin{abstract}
We review the theoretical description of spin-orbit scattering 
and electron spin resonance in carbon nanotubes. 
Particular emphasis is laid on the effects of electron-electron interactions.
The spin-orbit coupling is derived, and the 
resulting ESR spectrum is analyzed both using the effective low-energy field theory
and numerical studies of finite-size Hubbard chains and two-leg Hubbard ladders. 
For single-wall tubes, the field theoretical description predicts a
double peak spectrum linked to the existence of spin-charge separation.
The numerical analysis basically confirms this picture, but also predicts
additional features in finite-size samples.
\end{abstract}



\maketitle

\section{Introduction}

Nanotubes constitute a new class of mesoscopic quantum wires characterized
by the interplay of strong electron-electron interactions,
low dimensionality,  disorder, and unconventional spin dynamics
\cite{dekker99,bockrath,yao,postma,forro,egger97,egger2,balents}.  
In a sense, they represent an ideal model for strongly correlated
mesoscopic systems, where in fact basically all known 
effects in mesoscopic physics
have been experimentally observed. 
Two main classes of nanotubes can be distinguished,
namely single-wall nanotubes (SWNTs) which consist
of just one wrapped-up graphene sheet with radius $R$ in the 
nanometer regime, and
multiwall nanotubes which  contain additional inner shells \cite{forro}.
Here we will focus on the conceptually simplest case of {\sl metallic
SWNTs}, where interactions should
completely destroy the Fermi liquid picture and  imply
a so-called Luttinger liquid (LL) state of matter \cite{egger97,egger2}.
The Luttinger liquid is the generic low-energy description of metallic 1D 
(single-channel) quantum wires \cite{GNT}. 
 
Evidence for the LL behaviour of interacting
1D electrons has been reported for charge transport 
in SWNTs \cite{bockrath,yao,postma}.  However, in such materials one also
expects to find more dramatic consequences of the breakdown
of Fermi liquid theory, most notably the phenomenon of
 spin-charge separation.  This many-body effect asserts that electrons
brought into a LL effectively break up into a charge and a spin part
that travel with different velocities and hence will be spatially
separated after some time.  A recent proposal to detect evidence for
spin-charge separation in SWNTs has been based on spin transport \cite{balents}.
A different (and perhaps easier to realize) proposal based
on electron spin resonance (ESR) is reviewed in this paper,
expanding on our short paper \cite{martino}.
ESR is a valuable experimental tool
to probe the intrinsic spin dynamics of many systems.
In ESR experiments one applies a static magnetic field
and measures the absorption of radiation polarized
perpendicular to the field direction. In the absence of 
$SU(2)$ spin symmetry breaking terms in the Hamiltonian,
the absorption intensity is then simply a $\delta$-peak 
at the Zeeman energy \cite{OA,OA2}.

Since spin-orbit (SO) interactions are generally the
leading terms breaking the $SU(2)$ invariance, 
deviations in the ESR intensity from the $\delta$-peak, e.g.~shifts or
broadenings, are directly connected to these couplings.
Below we theoretically address the spin-orbit interaction and the
resulting ESR spectrum for interacting SWNTs, using both a continuum
field theory and a Hubbard model description. 
Within the effective field theory, 
the single $\delta$-peak is split
into {\sl two} narrow peaks in SWNTs  if spin-charge separation is realized.
Otherwise the ESR spectrum would form
a broad band with thresholds at the lower and upper edge 
 \cite{esrqw}.  This qualitative difference
is caused by the fact that the SO interaction in SWNTs
does not spoil spin-charge separation to leading order.
Experimental observation of the peak splitting could
therefore provide evidence for the elusive
phenomenon of spin-charge separation \cite{GNT}.
To experimentally check the predictions made below,
 samples free of magnetic impurities have to be used.
Such impurities have probably spoiled previous 
ESR measurements for nanotubes \cite{forro}.

The outline of this article is as follows. In Section \ref{sec2}
the Luttinger liquid theory of carbon nanotubes is reviewed in detail, and  we 
give an introduction to ESR theory as relevant for our purposes.  
The spin-orbit interaction is derived in Section \ref{sec3},
followed by a detailed discussion of the low-energy theory predictions for
the ESR spectrum in Section \ref{sec4}.  
An alternative approach is to use numerical methods to compute the
ESR spectrum for microscopic lattice fermion models. We shall use
both a Hubbard chain and the more realistic
two-leg Hubbard ladder formulation of interacting SWNTs
\cite{balents96}.
While charge transport does not allow for such a description due
to the importance of long-range interactions, it turns out that for
ESR spectra, only short-range interactions are important. These
are correctly captured by Hubbard-type models, and therefore
such models are expected to be  appropriate for the 
quantitative description of ESR spectra in SWNTs.
We discuss this approach in Section \ref{sec5} and compare
the numerical results with the field-theoretical predictions.
Finally, in Section \ref{sec6} conclusions and a brief outlook are
provided.  In most of the paper, we use the conventions $\hbar=c=1$ to 
simplify notation.

\section{Basics}
\label{sec2}

\subsection{Luttinger liquid theory}

Starting from a microscopic lattice description of the SWNT,
inclusion of the 
interactions among electrons  leads to rather complicated
models.  In the case of short-ranged interactions, one can
study Hubbard-type models, and we will do so in Section \ref{sec5}.
For low-energy phenomena such as
electron spin resonance, however, only bands close to the Fermi energy
do matter.  The relevant electronic properties of (not too thin) SWNTs are then 
caused only by the $\pi$ electrons of the wrapped graphene sheet.
On energy scales $|E| <D=\hbar v_F/R \approx 1$~eV 
 around the Fermi energy $E_F$ (here $v_F$ is the Fermi velocity),  the 
graphene bandstructure takes a simple form allowing to develop
a powerful field-theoretic framework for SWNTs \cite{egger97} reviewed below. 
We mention in passing
that it can be explicitly demonstrated that bands sufficiently far away
from the Fermi surface will not change the results obtained from
the field theory; for an explicit discussion, see Ref.~\cite{GNT}.

Simple tight-binding bandstructure calculations for graphene reveal that
there are only two linearly independent Fermi points (``flavours'')
with coordinates
$\alpha \vec{K}$ in the first Brillouin zone   ($\alpha=\pm$), 
instead of a continuous Fermi surface \cite{dekker99}. 
For $|E| < D$, the dispersion relation 
around the Fermi points is highly linear (two-dimensional light cone).
Since the basis of the graphene honeycomb lattice contains two atoms,
there are two sublattices $p=\pm$, and hence  
two degenerate Bloch states 
\begin{equation}\label{bloch}
\varphi_{p\alpha}(\vec{r}) =
(2\pi R)^{-1/2} \exp( -i\alpha \vec{K}\vec{r} ) 
\end{equation}
at each Fermi point $\alpha=\pm$.
Here  $\vec{r}=(x,y)$  lives
on the  sublattice $p$ under consideration, and
the correct normalization for nanotubes has been anticipated.
The Bloch functions are defined separately on each
sublattice such that they vanish on the other.
One can then expand the electron operator
in terms of these Bloch functions.
The resulting effective low-energy theory of
graphene is the 2D massless Dirac Hamiltonian,
as follows also from standard  $\vec{k}\cdot \vec{p}$ theory.

Wrapping the graphene sheet onto a cylinder then leads
to transverse momentum quantization, and hence to the 
effectively 1D bandstructure of a metallic SWNT.
Taking the $x$-axis along the tube direction and 
the circumferential variable as $0< y <2\pi R$,
quantization of transverse motion now allows for a
contribution $\propto \exp(i m y/R)$ to the wavefunction. 
However,  excitation of angular momentum states
other than $m=0$ costs a huge energy of the order $D$.
 Assuming that the SWNT is not
excessively doped, following our above remarks, in the field theory 
we may then omit all transport bands except $m=0$.
The theory will then apply on energy scales $|E|<D$ and 
lengthscales larger than the graphene lattice spacing $a\approx 0.246$~nm.

Evidently, a SWNT then forms a truly 1D quantum wire 
with only two spin-degenerate bands
intersecting the Fermi energy.  To take this into account,
the electron operator for spin $\sigma=\pm$
is written as
\begin{equation}\label{expa}
\Psi_\sigma(x,y) = \sum_{p\alpha} \varphi_{p\alpha}(x,y) 
\,\psi_{p\alpha\sigma} (x) ,
\end{equation} 
which introduces 1D fermion operators $\psi_{p\alpha\sigma}(x)$.
Neglecting interactions for the moment, 
the Hamiltonian is
\begin{equation} \label{h0}
H_0= -  v_F \sum_{p\alpha\sigma} p 
\int dx \;\psi_{p\alpha\sigma}^\dagger
 \partial_x \psi^{}_{-p\alpha\sigma},
\end{equation}
which is equivalent to a massless 1D Dirac Hamiltonian, where  
$v_F=8\times 10^5$~m/sec.

Next we discuss interactions mediated by the (possibly 
externally screened) Coulomb potential $U(\vec{r}-\vec{r}')$. 
The precise form of this potential will of course depend on details 
of the setup.  In the simplest case, bound electrons 
and the effects of an insulating substrate
are described by a dielectric constant 
$\kappa$,  and for the long-range Coulomb interaction, 
\begin{equation}\label{unsc}
U(\vec{r}-\vec{r}') = \frac{e^2/ \kappa} 
{\sqrt{ (x-x')^2 + 4R^2 \sin^2[(y-y')/2R] + a_z^2 }},
\end{equation}
where $a_z\approx a$ denotes the average distance between a
$2p_z$ electron and the nucleus, 
i.e.~the ``thickness'' of the graphene sheet.
Electron-electron interactions
are then described by 
\begin{equation}\label{int0}
H_I = \frac12 \sum_{\sigma\sigma'}\int d\vec{r} \int
d\vec{r}' \, \Psi^\dagger_\sigma(\vec{r}) \Psi^\dagger_{\sigma'}
(\vec{r}') 
 U(\vec{r}-\vec{r}') \Psi^{}_{\sigma'}(\vec{r}')
\Psi^{}_\sigma(\vec{r}) 
\end{equation}
which is brought into a 1D form by inserting equation  (\ref{expa}) 
for the electron field operator, allowing to employ 
the large arsenal of theoretical
methods available only in 1D \cite{GNT}.
The result is
\begin{eqnarray}\label{general}
H_I&=& \frac12 \sum_{pp'\sigma\sigma'} 
\sum_{\{\alpha_i\}}
 \int dx dx'\; V^{pp'}_{\{\alpha_i\}}(x-x')  \\ \nonumber
&\times&
\psi^\dagger_{p\alpha_1\sigma}(x) \psi^\dagger_{p'\alpha_2\sigma'}
(x') \psi^{}_{p'\alpha_3\sigma'}(x') \psi^{}_{p\alpha_4\sigma}(x) 
,
\end{eqnarray} 
with 1D interaction potentials
\begin{equation}\label{intpot}
V^{pp'}_{\{\alpha_i\}}(x-x') = \int dy dy'  
\varphi^{\ast}_{p\alpha_1}(\vec{r})
\varphi^{\ast}_{p'\alpha_2}(\vec{r}') 
U(\vec{r}-\vec{r}' + p \vec{d} \delta_{p,-p'} ) 
\varphi^{}_{p'\alpha_3}(\vec{r}') \varphi^{}_{p\alpha_4}(\vec{r}) .
\end{equation} 
These potentials only depend on $x-x'$ and on the 1D
fermion quantum numbers.
For interactions involving different 
sublattices $p\neq p'$ for $\vec{r}$
and $\vec{r}'$ in equation (\ref{int0}), 
a sublattice shift vector $\vec{d}$ arises \cite{egger2}.
To simplify the resulting 1D interaction (\ref{general}), 
one can exploit momentum conservation. Provided  we
stay away from the charge neutrality point $E_F=0$,
Umklapp electron-electron scattering processes can be ignored, and
the situation simplifies considerably.
We then have  only ``forward scattering'' processes \cite{egger2},
where $\alpha_1=\alpha_4$ and $\alpha_2=\alpha_3$, and
``backscattering'' processes
 with $\alpha_1=-\alpha_2=
\alpha_3=-\alpha_4$.

Next we introduce the potential
\begin{equation}\label{v0}
V_0(x-x')= \int_0^{2\pi R} \frac{dy}{2\pi R}
\int_0^{2\pi R} \frac{dy'}{2\pi R}\;  U(\vec{r}-\vec{r}') .
\end{equation}
For the unscreened Coulomb interaction  (\ref{unsc}), this
can be explicitly evaluated \cite{egger2}. 
$ $From equations (\ref{intpot}) and (\ref{bloch}),
the forward scattering  interaction potential reads
$V_0(x)+\delta_{p,-p'} \delta V_p(x)$,
with 
\begin{equation}\label{deltav}
 \delta V_p(x)  =  \int_0^{2\pi R} \frac{dy dy'}{(2\pi R)^2}
 [U(x+pd_x,y-y'+p d_y)- U(x,y-y') ] ,
\end{equation}
which is only present if $\vec{r}$ and $\vec{r}'$ are
located on different sublattices.  Thereby 
information about the discrete nature of the graphene network
has been kept despite the low-energy continuum approximation.
Since  $V_0(x)$ treats both sublattices on equal footing,
the resulting forward scattering interaction part couples
 only the total 1D electron densities, 
\begin{equation}\label{fs0}
H_I^{(0)} = \frac12 \int dx dx' \, \rho(x) V_0(x-x') \rho(x'),
\end{equation}
where 
$\rho(x) = \sum_{p\alpha\sigma} \psi^\dagger_{p\alpha\sigma}
\psi^{}_{p\alpha\sigma}$.  This part of the electron-electron
interaction is the most important one and is responsible for the 
LL behavior.  Note that it is due to the
{\sl long-ranged}\, tail of the Coulomb interaction. 
The remaining interactions originate from short-ranged 
interaction processes, and since these are effectively averaged
over the tube circumference, their amplitude is quite
small, scaling as $1/R$. Such couplings are 
seen below to cause only exponentially small gaps.

For $|x|\gg a$,  the term $\delta V_p(x)$ is extremely small.
However, for $|x|\leq a$, an additional term beyond equation (\ref{fs0}) arises
due to the hard core of the Coulomb interaction. 
At such small length scales, the difference between
inter- and intra-sublattice interactions matters, and
  $\delta V_p(0)$  must be computed
microscopically, leading to a coupling constant $f$
characterizing the additional forward scattering contribution 
\begin{equation}\label{fs1}
H^{(1)}_I= - f \int dx\sum_{p\alpha\alpha'\sigma\sigma'}
\psi^\dagger_{p\alpha\sigma}\psi^\dagger_{-p\alpha'\sigma'}
\psi^{}_{-p\alpha'\sigma'} \psi^{}_{p\alpha\sigma} ,
\end{equation}
where $f/a = \gamma_f e^2/R$ with a dimensionless constant
$\gamma_f$ depending on the tube chirality.  
An estimate  for armchair SWNTs 
yields $\gamma_f\approx 0.05$, implying that $f$ is 
 very small.
A similar reasoning applies to the backscattering contributions 
 $\alpha_1=-\alpha_2=\alpha_3=-\alpha_4$
in equation \ref{general}). 
Because of a rapidly oscillating phase factor, 
again the only contribution
comes from $|x-x'|\leq a$.
Furthermore, only the
part of the interaction which does not distinguish among
the sublattices is relevant, leading to
\begin{equation} \label{bs}
H_I^{(2)} = b \int dx\sum_{pp'\alpha\sigma\sigma'}
\psi^\dagger_{p\alpha\sigma}\psi^\dagger_{p'-\alpha\sigma'}
\psi^{}_{p'\alpha\sigma'} \psi^{}_{p-\alpha\sigma} .
\end{equation}
For the unscreened interaction (\ref{unsc}), 
$b/a = \gamma_b e^2/R$ with $\gamma_b\approx \gamma_f$.
For externally screened Coulomb interaction, however, one may have
$b\gg f$.

Further progress can be made by using the Abelian bosonization
approach \cite{GNT}.
For that purpose, one brings the non-interacting
Hamiltonian (\ref{h0}) into the conventional form
of the 1D Dirac model by
switching to right- and left-movers
($r=\pm$) which are linear combinations of the sublattice
states $p=\pm$.
In this representation, an Abelian
 bosonization formula  \cite{egger97,egger2,GNT}
applies  with four bosonic phase fields 
$\theta_a(x)$ and their canonical momenta $\Pi_a(x)$.
The four channels are obtained from
combining charge and spin degrees 
of freedom as well as symmetric and antisymmetric linear
combinations of the two Fermi points, $a=c+,c-,s+,s-$.
The bosonized expressions for $H_0$ and $H_I^{(0)}$  read 
\begin{eqnarray} \label{bh0}
 H_0 &=& \sum_{a} \frac{v_F}{2} 
\int dx \left[  \Pi_a^2
+ K^{-2}_a (\partial_x \theta_a)^2 \right]\\
\label{bfs0}
H_I^{(0)} &=& \frac{2}{\pi} \int dx dx' \;
\partial_x\theta_{c+}(x) V_0(x-x') \partial_{x'} \theta_{c+}(x') .
\end{eqnarray}
The bosonized form of $H_I^{(1,2)}$ \cite{egger97} leads to nonlinearities
in the $\theta_a$ fields for $a\neq c+$. 
Although bosonization of equation (\ref{h0}) gives $K_a=1$ in
equation (\ref{bh0}), interactions will renormalize these parameters.
In particular, in the long-wavelength limit,
 $H_I^{(0)}$ can be incorporated into $H_0$ by putting 
\begin{equation}\label{Kdef}
K_{c+} = \left \{1+ 4\widetilde{V}_0(k\simeq 0)/\pi \hbar v_F \right\}^{-1/2}
 \leq 1 ,
\end{equation}
while for all other channels, the coupling
constant $f$  gives rise to a tiny renormalization,
$K_{a\neq c+}= 1+  f /\pi \hbar v_F \simeq 1$. 
The plasmon velocities of the four modes are
$v_a=v_F/K_a$, and hence the charged $(c+)$ mode 
propagates with higher velocity than the three neutral modes. 
The dimensionless Luttinger parameters measure the correlation strength
in the system, with the noninteracting point at $K_a=1$ and repulsive
interactions leading to $K_{c+} <1$.
For the long-ranged interaction (\ref{unsc}), 
the logarithmic singularity in $\widetilde{V}_0(k)$
requires the infrared cutoff $k=2\pi/L$ due to the finite
length $L$ of the SWNT, resulting in:
\begin{equation} \label{longr}
 K_{c+}= \left\{ 1+\frac{8e^2}{\pi\kappa\hbar v_F} \ln(L/2\pi R) 
\right\}^{-1/2} .
\end{equation}
Since $\hbar c/e^2\simeq 137$, 
we  estimate
$e^2/\hbar v_F = (e^2/\hbar c) (c/v_F) \approx 2.7$, and
therefore $K_{c+}$ is typically in the range $0.2$ to $0.3$.
This estimate does only logarithmically depend on $L$ and $R$, 
and therefore is expected to be almost independent of the 
 sample under study.  The Luttinger parameter (\ref{longr}) 
can also be written in the form
\[
K_{c+} = \left(1 + \frac{2E_c}{\Delta} \right)^{-\frac{1}{2}} ,
\]    
where $E_c$ is the charging energy and
$\Delta$ the single-particle level spacing.
The small value predicted here
implies that a metallic SWNT should be a strongly correlated system
displaying pronounced non-Fermi liquid effects.

It is clear from equations (\ref{bh0}) and (\ref{bfs0}) that
for $f=b=0$, a SWNT constitutes a realization of
the LL. We therefore have to address the
effect of the nonlinear terms associated with the
coupling constants $f$ and $b$.  This can be done
by means of the renormalization group approach.
Together with a  solution via Majorana
refermionization,  this procedure allows for the complete
characterization of the non-Fermi-liquid ground state of a clean
nanotube \cite{egger97,egger2}.  $ $From this analysis, one finds that for 
temperatures above the exponentially small energy gap 
\begin{equation}\label{massb}
k_B T_b = D \exp[-\pi \hbar v_F / \sqrt{2} b] 
\end{equation}
induced by backscattering processes,
the SWNT is adequately described by the LL model, and $H_I^{(1,2)}$ 
can effectively be neglected.  A rough order-of-magnitude
estimate is $T_b \approx 0.1$ mK.  In the remainder, we focus on
temperatures well above $T_b$, where the nonlinearities can be 
neglected and the Luttinger picture applies.

\subsection{Sugawara formulation}

In the following discussion of ESR theory, it is mandatory to  
keep the $SU(2)$ spin symmetry explicit at all stages.  
To do so it is advantageous to avoid the
Abelian bosonization used above, which breaks the spin symmetry by hand,
but rather employ the Sugawara 
formulation \cite{GNT} which manifestly respects $SU(2)$ spin invariance. 
This formulation is in fact fully equivalent to a Wess-Zumino-Witten
theory, even with the flavour degeneracy due to the two
Fermi points \cite{GNT}.   
Since our main interest is on spin properties, we shall however
suppress the flavour index for most of what follows, 
but return to the complexities added by 
it later, see Section \ref{sec5}.  
Instead of four channels, we then have a two-channel Luttinger liquid,
with interaction constants for charge, $K_c=(K_{c-}+K_{c+})/2\simeq
(1+K_{c+})/2$, and
spin, $K_s=(K_{s-}+K_{s+})/2\simeq 1$.
In Abelian bosonization, in order
 to impose the correct spin symmetry, one then fixes $K_s=1$ \cite{GNT}.
This procedure is not necessary in the Sugawara treatment below.
Inclusion of  magnetic Zeeman fields $B$
(orbital effects play no role here) only affects 
the spin sector and in general could renormalize $K_s$.
However, this renormalization is irrelevant to ESR, 
which probes the finite energy scale $\approx B$ \cite{OA}.
 In addition, the velocities $v_{c/s}=v_F/K_{c/s}$ are
used in the following.  

The Sugawara formulation uses the charge ($J_{L/R}$) and spin currents
$\vec J_{R/L}$ for right-  and left- ($r=\pm=R/L$) moving
electrons described by chiral 1D fermion operators $\psi_{r=R/L}(x,t)$,
where we suppress the additional spin index $\sigma$.
In  terms of these
field operators, chiral $SU(2)$ spin current operators are given by 
\begin{equation} \label{spincurr}
\vec J_{R,L}(x) = \frac12 : 
\psi_{R/L}^\dagger(x) \vec\sigma \psi_{R/L}^{}(x) : ,
\end{equation} 
where the colons denote normal-ordering and Pauli matrices 
$\vec \sigma$ act in spin space.
They obey Kac-Moody commutation relations ($\mu,\nu=x,y,z$) 
\begin{equation}
[J^\mu_{L/R}(x),J^\nu_{L/R}(x')]= \pm i \delta'(x-x')\delta^{\mu\nu}/4\pi
+i\epsilon^{\mu\nu\lambda} J^\lambda_{L/R}(x) \delta(x-x').
\end{equation}
Likewise, charge current operators are defined as
\begin{equation} \label{charcurr}
 J_{R,L} = : \psi_{R/L}^\dagger(x) \psi_{R/L}^{}(x) : ,
\end{equation} 
where spin indices are summed over.
Using these current operators, the
Luttinger liquid in the $U(1)\times SU(2)$ invariant
Sugawara formulation reads \cite{GNT},
\begin{equation}\label{hamll}
H_0= H_c + H_s,
\end{equation}
with decoupled charge and spin parts. The charge sector is described by
the $U(1)$ invariant Hamiltonian
\begin{equation}\label{chargepart}
H_c =  \frac{\pi v_c}{2} \int dx ( : J_R J_R + J_L J_L : + 
g_c : J_L J_R: ).
\end{equation}
The coupling $g_c$ is determined
by the Luttinger parameter $K_c$. Explicit expressions can be found in
Ref.~\cite{GNT}, but they are not required below.  In particular,
in the noninteracting
limit, $g_c=0$ and $v_c=v_F$.
The $SU(2)$ invariant spin Hamiltonian $H_s$ commuting with $H_c$ is
\begin{equation}\label{spinpart}
H_s= \frac{2\pi v_s}{3} 
\int dx : \vec J_R \cdot \vec J_R+ \vec J_L\cdot \vec J_L :
+ g_s \int dx : \vec J_R \cdot \vec J_L :,
\end{equation}
where $v_s\approx v_F$ is the spin velocity
and $g_s=b v_s$ is the coupling constant for
electron-electron backscattering processes.
In a dynamical spin-sensitive ESR measurement, one should not
simply discard this coupling despite the above thermodynamic argument
invoking the smallness of gaps \cite{balents}.
The LL Hamiltonian (\ref{hamll}) completely decouples when expressed 
in terms of spin and charge currents (or the spin/charge bosons
of Abelian bosonization).  This remarkable fact leads to the
phenomenon of spin-charge separation, and  
unless the spin-orbit interaction couples spin and charge sectors,
ESR will only probe the spin sector.

\subsection{Electron spin resonance}

Adopting the conventional Faraday configuration,
the ESR intensity at frequency $\omega$ is 
proportional to the Fourier transform of the transverse spin-spin 
correlation function \cite{OA},
\begin{equation} \label{int1}
I(\omega)= \int \! dt \, e^{i\omega t} \langle S^+(t) S^-(0) \rangle ,
\end{equation}
where the static magnetic field points along the $z$-axis,
$\vec S = \sum_i\vec S_i $ is the total spin operator, and
$S^{\pm} = S^x \pm i S^y$. 
The Hamiltonian can be written as  $H=H_0 + H_Z + H'$,
where $H_0$ represents the $SU(2)$ invariant nanotube model (\ref{hamll})
including electron-electron interactions, 
$H_{Z} = - g_e \mu_B \vec B \cdot\vec S$ is the Zeeman term
(below, often $g_e \mu_B =1$),
and $H'$ represents
$SU(2)$ spin-symmetry breaking terms, in particular the SO coupling. 
Inserting a complete set of eigenstates $|a \rangle$ of $H$ in 
equation (\ref{int1}), the ESR intensity  follows as
\begin{equation}
I(\omega) = \frac{1}{Z} \sum_{a,b} e^{- E_b/k_B T} 
\delta(\omega -(E_a-E_b)) \,
| \langle a | S^- | b \rangle |^2.
\end{equation}
In the absence of $H'$, there are only 
contributions from matrix elements between 
eigenstates with equal total spin $S_a= S_b$.  Then 
all states with $S^z_a=S^z_b-1$ will lead to a 
$\delta$-peak at frequency $\omega=B$.
For instance, at zero temperature,  the application of a magnetic field $B$, 
taken as large enough to overcome a spin gap possibly present at $B=0$, 
leads to a ground state with finite magnetization, $S_0 \neq 0$, and 
the states with $S^z_a=S^z_0-1$ again yield the
$\delta$-peak.
This can be made explicit as follows. Since 
$\left[ H_0 + H_{Z}\, , S^-\right] = BS^- $, one has 
$(H_0+H_{Z}) S^-|0\rangle = S^-(H_0+H_{Z}) |0\rangle + BS^-|0\rangle$ 
and identifying $S^-|0\rangle = |a\rangle$, one gets  
$E_a= E_0 + B$ and thus
$I(\omega)= I_0 \delta(\omega-B)$. Any perturbation
 preserving $SU(2)$ invariance 
will neither shift nor broaden this peak, even
at finite temperature \cite{OA2}.
To get nontrivial ESR spectra, one has to 
identify the leading perturbation breaking 
$SU(2)$ invariance, which will cause finite linewidth and shift of 
the ESR peak.
In metallic systems like SWNTs, one has to consider the 
spin-orbit (SO) interaction. This is done in the next section.

\section{Spin-Orbit Coupling}
\label{sec3}

\subsection{Microscopic derivation}

In our derivation of the SO term, we shall neglect electron-electron 
interactions.  Local electric fields exerted by other electrons on a given
electron are typically weak compared to the ionic fields \cite{chenraikh},
and will generally only weakly renormalize the SO couplings from
their noninteracting values.
In a single-particle picture, the SO interaction then appears because an 
electron moving in the electrostatic potential
$\Phi(\vec r)$ experiences an effective magnetic field 
$\vec v \times \nabla \Phi$ in its rest frame.
 With $\vec p = m \vec v$ the SO interaction
reads in second-quantized form
\begin{equation} \label{SOham2}
H' = -\frac{g_e \mu_B}{4m} \int d\vec{r} \, \, \Psi^\dagger \left[ 
(\vec p \times \nabla \Phi) \cdot \vec \sigma \right] \Psi.
\end{equation}
This represents the starting point for our discussion of the SO coupling.

For a microscopic lattice description,
the electron field operator
$\Psi_\sigma(\vec r)$ [whose low-energy expansion is given above in
equation (\ref{expa})]
can now be expressed in terms of the electron operators $c_i$ 
for honeycomb lattice site $i$ at $\vec r_i$,
\begin{equation}\label{expa1}
\Psi_\sigma (\vec r) = \sum_i \chi(\vec r - \vec r_i) \ c_{i\sigma},
\end{equation}
where $\chi(\vec r-\vec r_i)$
 is the corresponding 
Wannier wavefunction centered at lattice site $\vec r_i$.  
These localized Wannier orbitals can be chosen 
as real-valued functions even when hybridization 
with $s$-orbitals is important.
For the simplest case,  $2p_z$ orbital
wavefunctions could be used,
with $z$ perpendicular to the graphene plane,
\[
\chi(\vec r) =(z/4a_0\sqrt{2\pi}) \exp(- r/2a_0),
\]
 where 
$a_0 = \hbar^2/6me^2$ is the effective Bohr radius. 

Inserting the expansion (\ref{expa1}) into (\ref{SOham2}), 
 the SO interaction reads \cite{martino,bonesteel,ando},
\begin{equation} \label{SOham3}
H' = \sum_{\langle jk \rangle} \,\,
 i c^\dagger_j (\vec u_{jk} \cdot \vec \sigma) c^{}_k  + \,{\rm H.c.}  
\end{equation}
which explicitly breaks $SU(2)$ symmetry.
The SO vector $\vec u_{jk} = -\vec u_{kj}$ has real-valued entries
and can be written as
\begin{equation}\label{sovect}
\vec u_{jk} = \frac{g_e\mu_B}{4m} \int d\vec r \, \Phi(\vec r)
\left[ \nabla \chi (\vec r - \vec r_j) \times 
\nabla \chi (\vec r - \vec r_k)  \right]  .
\end{equation} 
The on-site term ($j=k$) is identically zero, and
since the overlap decreases exponentially with $|\vec r_j - \vec r_k|$, 
we keep only nearest-neighbour terms in equation (\ref{SOham3}). 
Equation (\ref{SOham3}) also allows to incorporate
electric fields due to impurities or close-by gate electrodes.
To connect the SO vector to the experimentally measurable
SO relaxation rate, $\tau_{SO}^{-1}$, we estimate this rate
using Fermi's golden rule. Assuming a constant $\vec u$, 
the probability for a transition from the initial state 
$|i\rangle = |k, \uparrow\rangle$ to the final state 
$|f\rangle = |k', \downarrow\rangle$ is given by
$\Gamma_{i \rightarrow f} = 2\pi |\langle f| H'| i \rangle |^2 
\delta(E_i-E_f)$, and 
summing over all final states,
\begin{equation}\label{SOrate}
\tau_{SO}^{-1} = 8  (u^2_x + u^2_y)/3 v_s.
\end{equation}
For SWNTs, SO couplings are generally expected to be small.
This is in accordance with our approach
since the SO vector (\ref{sovect}) vanishes by symmetry
for an ideal 2D honeycomb lattice.  A finite nearest-neighbour
SO coupling can only arise due to the finite curvature,
stray fields from nearby gates, or due to impurities, 
which break the high symmetry.
Focusing for clarity on the curvature-induced SO coupling,
the SO vector for achiral tubes can be seen to only depend on bond direction,
\begin{equation}\label{uas}
\vec u_{\vec r_i, \vec r_i+\vec \delta_n} = \vec u_n,
\end{equation}
where the nearest-neighbour bonds $\vec\delta_n$ ($n=1,2,3$) connect
the two sublattices of the graphene sheet in real space. 
For explicit representations of the 
vectors $\vec\delta_n$, see e.g. \cite{ando}. 
 
\subsection{Low-energy form}

{}From now on we then restrict our analysis to achiral tubes and
construct the low-energy field theory Hamiltonian $H'$ describing the
spin-orbit interaction (\ref{SOham3}), including the 
flavour indices for the moment. Using equation (\ref{uas}), the
microscopic form is 
\[
H'= \sum_{\vec r} \sum_{n=1}^3 \left( i c^\dagger_{\vec r,p=+}
 \vec u_n\cdot
\vec \sigma c_{\vec r+\delta_n,p=-}^{} + {\rm H.c.} \right), 
\]
where $\vec r$ runs over all lattice sites of the sublattice $p=+$, and
we have made the sublattice dependence of the lattice
 electron operators explicit
(again spin indices are implicit).
The expansion (\ref{expa}) implies
\[
c_{\vec r,p,\sigma} =  (2\pi R)^{-1/2}\sum_{\alpha} e^{-i\alpha \vec K\cdot
\vec r} \psi_{p\alpha\sigma}(x),
\]
which yields after $y$-integration and Taylor  expansion of 
$\psi_{p=-,\alpha\sigma}(x+\hat e_x\cdot \vec\delta_n)$ the result
\begin{eqnarray}\label{SOham4}
H'&=& \int dx \sum_{\alpha,n} e^{-i\alpha \vec K \cdot \vec\delta_n}
\psi^\dagger_{p=+,\alpha}(x) \left(i\vec u_n\cdot \vec \sigma\right) \\
&\times& \nonumber
\left[ \psi_{p=-,\alpha}(x)+ (\hat e_x\cdot \vec\delta_n) \partial_x
\psi_{p=-,\alpha}(x) \right] + {\rm H.c.}
\end{eqnarray}
where we use the unit vector $\hat{e}_x$  
along the tube axis. Terms involving higher order derivatives of
the field operators are highly irrelevant and can safely be ignored.
In equation (\ref{SOham4}), we have also neglected oscillatory terms that vanish 
in doped SWNTs because of axial
momentum conservation.  
These oscillations are governed by the wavevector $2|\mu|/v_F$ 
corresponding to the doping level $\mu$. 
In practice, intrinsic doping is unavoidable, but for 
a commensurate situation ($\mu=0$), additional terms do arise that
are ignored in this section.

Next we define vectors given in terms
of the vectors specified in (\ref{uas}),
\begin{eqnarray*}
\vec u_{1\alpha} &=& \sum_{n=1}^3 \exp(-i\alpha\vec K\cdot\vec
 \delta_n) \vec u_n\\
\vec u_{2\alpha} &=& \sum_{n=1}^3 \exp(-i\alpha\vec K\cdot\vec \delta_n) 
\vec u_n (\hat e_x\cdot \vec \delta_n).
\end{eqnarray*}
We then obtain from equation (\ref{SOham4}) the form $H'=H_1+H_2$ with
\begin{equation}\label{SOham5a}
H_1= \int dx \sum_{\alpha} 
\psi^\dagger_{p=+,\alpha}(x) \left(i\vec u_{1,\alpha}
\cdot \vec \sigma\right) 
\psi_{p=-,\alpha}(x) + {\rm H.c.}
\end{equation}
and
\begin{equation}\label{SOham5b}
H_2= \int dx \sum_{\alpha} 
\psi^\dagger_{p=+,\alpha}(x) \left(i\vec u_{2,\alpha}
\cdot \vec \sigma\right) 
\partial_x \psi_{p=-,\alpha}(x) + {\rm H.c.}
\end{equation}
It is then apparent that $H_1$ will give the dominant contribution
(relevant in the renormalization group sense, scaling dimension 1), 
while $H_2$ is
marginal (scaling dimension 2). Since SO interactions are
always small, the renormalization group arguments around the 
Luttinger fixed point with no SO couplings can be safely used here.
Note that all contributions neglected here either violate
momentum conservation or have scaling dimensions $\geq 3$,
i.e.~are highly irrelevant.

The final step is to move from the sublattice
description to the $r=R/L$ basis using a unitary
transformation.  
Noting that the term involving ${\rm Re}\vec u_{1\alpha}$ involves
oscillatory contributions that can be neglected by virtue
of momentum conservation, the dominant low-energy SO contribution 
takes the form
\begin{equation} 
\label{contHam} 
H_1 = \int dx  \sum_\alpha \vec \lambda_\alpha \cdot ( \vec J_{L,\alpha}
 - \vec J_{R,\alpha} ),
\end{equation}
with effective SO vectors
\begin{equation}\label{las}
\vec \lambda_\alpha =  2\ \mbox{Im} \vec u_{1,\alpha} =
2\ \mbox{Im} \sum_n e^{-i\alpha\vec K \cdot \vec\delta_n} \vec u_n.
\end{equation} 
In the analysis of Sec.~\ref{sec4}, for clarity, we shall ignore the flavour
index, but it is clear from the above how to add it to the theory.
The subleading term $H_2$ reads (ignoring the flavour index,
and again omitting oscillatory terms)
\begin{equation}\label{contHam2}
 H_2 =   \int dx 
\sum_{r=R/L} \psi^\dagger_r \,\vec \lambda'\cdot \vec 
\sigma \, i \partial_x \psi_r  + \mbox{H.c.}
\end{equation}
where
\[
\vec \lambda' = \sum_n e^{-i\vec K \cdot\vec \delta_n}
(\hat{e}_x \cdot \vec \delta_n) \vec u_n .
\]

\section{ESR spectrum from field theory}
\label{sec4}

In the following, we will keep only the leading contribution
(\ref{contHam}) and neglect the marginal term $H_2$.
In the numerical analysis of Section \ref{sec5}, the full
SO interaction will be considered to assess the accuracy of
these approximations.  Since SO couplings are typically very
small,  however, the above perturbative reasoning, suggesting
to drop $H_2$ and the oscillatory terms, is expected to work.
Remarkably,  the SO interaction then acts exclusively 
in the spin sector and hence
does  not spoil spin-charge separation. 
Therefore the ESR intensity can be computed from the spin sector alone.

For convenience, we define
\begin{equation}\label{lamdef}
\vec\lambda_{r=R/L=\pm} = \vec B \pm \vec \lambda,
\end{equation}
which represents an effective magnetic field 
acting separately on the chiral (right/left-moving)
spin currents.  The presence of two pseudo-fields already hints at the
outcome of our calculation below, namely the emergence of a double peak spectrum.
Their position will then be given by the absolute values 
$\lambda_r=|\vec \lambda_r|$.  The spin Hamiltonian governing
the ESR spectrum $I(\omega)$ is then 
\begin{equation}\label{esrham1}
H = H_s - \sum_r \int dx\ \vec\lambda_r \cdot \vec J_r,
\end{equation}
with $H_s$ given in equation (\ref{spinpart}).
The Heisenberg equation of motion for the chiral spin 
current operators is
\begin{equation}\label{eom}
(\partial_t\pm v_s/3\partial_x) \vec J_r =\mp(g_s/4\pi)
\partial_x\vec J_{-r} + g_s \vec J_R\times \vec J_L -
\vec \lambda_r\times \vec J_r.
\end{equation}
Taking $\vec B=B \hat e_z$, and using $\vec S=\vec J_R
+ \vec J_L$, the ESR spectral density is
\begin{equation}\label{specdens1}
I(\omega) = \int dt dx e^{i\omega t} \sum_{rr'}
 \langle J^+_r(x,t) J^-_{r'}(0,0) \rangle.
\end{equation}
We then compute
the ESR spectrum using the Sugawara spin Hamiltonian (\ref{esrham1}). An 
independent route would proceed via fermionization \cite{martino}.
For simplicity, we discuss the case $g_s=0$ (no backscattering),
but the method is flexible enough to allow treatment of the general
case as well. However, the numerical analysis of Section \ref{sec5} indicates that 
indeed the backscattering interaction does not affect the results 
in a significant way.

Since the spin Hamiltonian decouples in the chiral spin currents,
we just need to add  the contributions due to the two chiralities.
In equilibrium, using the
spin susceptibility $1/4\pi v_s$ of a chiral fermion, we have
\begin{equation}
\langle \vec J_R\rangle = \frac{1}{4\pi v_s} \vec \lambda_R.
\end{equation} 
Using the fluctuation-dissipation theorem, we can now express
$I(\omega)$ in (\ref{specdens1}) in terms of the imaginary part
of the susceptibility tensor $\chi_{r}(q,\omega)$,
\begin{equation}\label{specdens2}
I(\omega)=\sum_{r=\pm}\sum_{\mu,\nu=x,y}
\frac{2}{1-\exp(-\omega/k_B T)} \, {\rm Im}
\chi_r^{\mu\nu}
(q\to 0,\omega).
\end{equation}
The susceptibility can in turn directly be obtained from the
equation of motion (\ref{eom}). Formally defining the matrix 
 $(\Lambda_r)^{\mu \nu} = \epsilon^{\mu\nu\alpha}\lambda^\alpha_r$, we find
\[
\chi_{r}(q\to 0,\omega) = \frac{1}{4\pi v_s} [i\omega + \Lambda_r]^{-1}
\Lambda_r,
\] 
and straightforward algebra then yields the ESR spectrum 
\begin{equation} \label{swnt}
I(\omega)=\sum_{r=\pm} \left[ 1+ \lambda^z_{r}/\lambda_r
\right]^2  \frac{\lambda_r}{4v_s(1-e^{-\lambda_r/k_B T})} 
\, \delta(\omega-\lambda_{r}) .
\end{equation}
As expected, the single $\delta$-peak  splits into two peaks
but there is no broadening.
The peak separation is $|\lambda_+ - \lambda_-|$, see (\ref{lamdef}),
 and
the peak heights are generally different.  To lowest order in
$\lambda/B$, the two peaks are located symmetrically around $\omega=B$.
Notice that for $\vec B$ perpendicular to the effective
 SO vector $\vec \lambda$, the splitting is zero.  

It should be stressed that these results hold both for
the non-interacting and the interacting case.  
However, for the interacting case realized in SWNTs
the double peak structure  is only possible if
spin-charge separation is present. Otherwise the
charge sector will mix in, leading to broad bands with 
threshold behaviours \cite{esrqw}.
Closer inspection shows that  inclusion of the subleading term 
(\ref{contHam2}) preserves the splitting into two peaks,
but the peaks now acquire a finite width $\sim |\vec \lambda'|$.
Similarly, the effects of backscattering are expected to be
small. 
It is important to point out that the double peak spectrum would
also show up in the non-interacting case, and hence its experimental
observation does not represent a true proof for spin-charge separation.
However, it would certainly give strong evidence, as in the interacting
situation the double peak survives only when spin-charge separation
is realized.

In practice, to get measurable intensities, one may have to 
work with an ensemble of SWNTs. The proposed experiment may be 
possible using electric-field-aligned SWNTs, or by employing 
arrays of identical SWNTS.
In more conventional samples with many SWNTs, however, the 
SO vector $\vec \lambda$ can take a random direction.
For results on the averaged spectrum, see \cite{martino}.

\section{ESR spectra for armchair SWNTs: Two-leg Hubbard ladders}
\label{sec5}

In this section we discuss an alternative
description of the ESR spectrum via a lattice fermion description. 
We shall numerically compute ESR spectra for a
Hubbard chain and a two-leg Hubbard ladder \cite{balents96}.
These models 
ignore the long-range part of the Coulomb interactions which makes
them less accurate for the description of charge transport.  Since the
long-range tails of the interaction do not affect 
the spin dynamics, however,
this does not create a problem for the spin problem at hand.

Modelling the SWNT by a two-leg Hubbard ladder  is attractive for
the following reasons:
\begin{itemize}
\item Backscattering
interactions as well as commensurability effects (Umklapp scattering)
are included.
\item  The two flavour degrees of freedom are taken care of.
\item Band curvature effects are taken into account.
\item The marginal SO term (\ref{contHam2}) as well as all irrelevant 
contributions can be kept.
\item For the Hubbard chain, it is known that any magnetic field
spoils spin-charge separation \cite{frahm}.  The importance of this
effect for SWNTs can be assessed.
\item Accurate numerical techniques are available for such models.
\end{itemize}
To gain a broader perspective and to also
make contact to other 1D conductors, before dealing with the more
complex two-leg ladder system,
we first consider a 1D Hubbard chain \cite{GNT} with uniform SO coupling,

\subsection{Hubbard chain}

The 1D Hubbard model  with a uniform SO coupling vector $\vec \lambda$ is defined by 
\begin{equation}  \label{lattHam}
H = \sum_{n} \left[c^\dagger_{n}
 \left(-t + i\vec{\lambda}\cdot\vec{\sigma} \right)  c_{n+1}
+ \,{\rm H.c.} \right]+ U \sum_n n_{n\uparrow} n_{n\downarrow}
 -B\sum_n S^z_n .
\end{equation}
For $\vec \lambda=0$, equation (\ref{lattHam}) leads to a LL phase at low 
energies \cite{GNT}.
It is worth noting that then the exact solution of (\ref{lattHam})
shows that with a magnetic field, the 
charge and spin sectors 
get mixed \cite{frahm}. 
To make progress, we gauge away the SO term by the unitary 
transformation
\[
c_n \rightarrow d_n = \left\{ \begin{array}{ll}
c_n &\quad \quad (n \; \; {\rm odd}) ~,\\
\exp (i \alpha \, \hat \lambda\cdot \vec \sigma) \, c_n & \quad \quad 
(n \; \; {\rm even}) ,
\end{array}
\right.
\]
where $\alpha = \arctan (\lambda/t)$, $\hat \lambda=
\vec \lambda/\lambda$ with
 $\lambda = |\vec \lambda|$. 
The hopping amplitude then changes as 
$t \rightarrow \tilde t = t + \lambda \sin \alpha$, 
and one can easily see that $n_{n\uparrow} n_{n\downarrow} =
\tilde n_{n\uparrow} \tilde n_{n\downarrow}$, where 
$\tilde n_{n\sigma} = d^\dagger_{n\sigma} d_{n\sigma}$. 
Then the Hamiltonian (\ref{lattHam}) can be rewritten as 
\[
H = -\tilde t \sum_{n} (d^\dagger_{n} d_{n+1} + 
\, {\rm H.c.} ) 
+ U \sum_n \tilde n_{n\uparrow} \tilde n_{n\downarrow} -B\sum_n \tilde 
S^z_n ,
\]
and the spin operator as
\[ 
\vec S_n = \left\{ \begin{array}{ll}
\frac12 d^\dagger_n \vec \sigma d_n & \quad \quad (n \;{\rm odd}) ~,\\
\frac12 d^\dagger_n ( e^{i\alpha \hat \lambda \cdot \vec \sigma} 
\vec \sigma e^{- i\alpha \hat \lambda\cdot \vec \sigma}) d_n & 
\quad \quad (n \; {\rm even}) .
\end{array}
\right.
\]
The transformed picture is easier to use in practical calculations.
We have calculated the ESR spectrum for (\ref{lattHam}) using exact
diagonalization for small lattices.  

\begin{figure} 
\begin{center}
\epsfysize=12cm
\epsffile{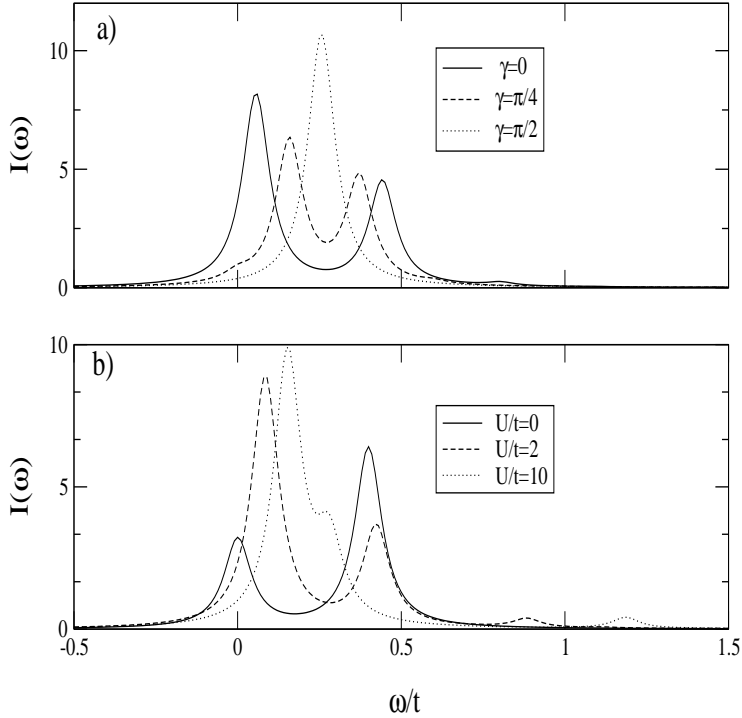}
\end{center} 
\caption{ \label{hubchain} 
Exact diagonalization results for the $T=0$ ESR spectra of a Hubbard chain
with  $\lambda=0.1$, $B /t=0.2$, and artificial 
broadening $0.05 t$.
In (a) ESR spectra for $U/t=1$ are shown for different angles $\gamma$ between 
$\vec{\lambda}$ and $\vec{B}$.  
In (b) the $U$-dependence is shown for $\gamma=0$.
All results are for 12 lattice sites and
4 electrons. }
\end{figure}

The main features emerging from exact 
diagonalization can be seen in figure \ref{hubchain}. 
For weak interactions, see figure \ref{hubchain}(a),
we recover the two-peak structure predicted by field theory, see (\ref{swnt}),
with the peaks symmetrically arranged around $\omega=B$ 
but with different peak heights.  In addition, 
no splitting is seen for $\vec B$ perpendicular to the SO vector 
$\vec \lambda$.  The interaction dependence of the ESR
spectrum is shown in figure \ref{hubchain}(b).
Strong interactions suppress one of the peaks 
and enhance the other, and also give rise to more structure in the spectrum.
This behaviour can be traced back to the effects of band curvature.
Note that for increasing $U$, the distance between both peaks
decreases.  This feature can be qualitatively explained
by a decrease of the spin velocity $v_s$ with $U$, {\it i.e. } for a flatter 
band, the spin-orbit splitting is less effective.

\subsection{Two-leg Hubbard ladder: Armchair SWNTs}

\begin{figure}
\begin{center}
\epsfysize=6cm
\epsfbox{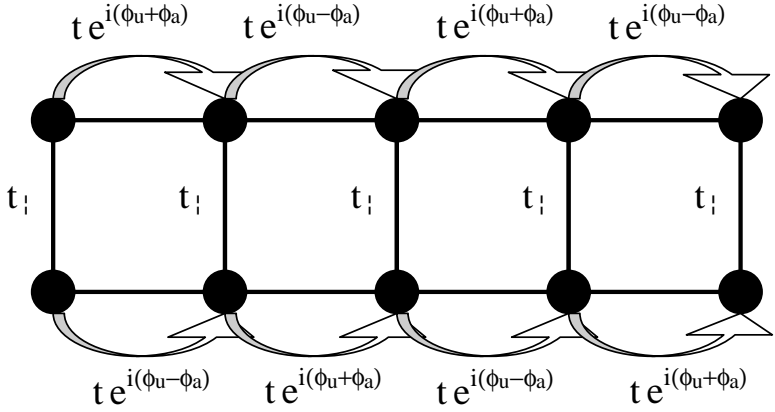} 
\end{center}
\caption{\label{ladderfas}
Sketch of the two-leg Hubbard ladder including spin-orbit couplings
of the type considered here. The arrows indicate the 
phase for spins up. Spins down have the opposite phase.}
\end{figure}

We are now ready to generalize the above treatment to armchair SWNTs.
Mapping the interacting honeycomb lattice Hamiltonian for armchair SWNTs
onto the two-leg Hubbard model \cite{balents96}, one finds
\begin{equation}
\label{eqnhubb}
H_0 =\sum_{ni}(-t c_{ni}^{\dagger} c^{}_{(n+1)i} 
 -  t_{\perp}  c_{n1}^{\dagger} c^{}_{n2}
+ {\rm H.c.}) +U\sum_{n,i} n_{n,i,\uparrow} n_{n,i,\downarrow } .
\end{equation}
While under the mapping,
 $t_\perp = t$ up to $1/R$ corrections \cite{balents96}, 
we shall allow for $t_\perp\neq t$ to
have better numerical accuracy.
Here we also  want to include the
spin-orbit interaction (\ref{SOham3}) in this mapping.
We use the SO vectors (\ref{uas}), and for simplicity 
set $\vec u_{n=1}=0$, since this SO vector does not appear in the
leading low-energy SO vector $\vec \lambda$ in (\ref{las}) for
armchair SWNTs. This simplification also allows for numerically feasible 
calculations.
Writing $\vec{w}_n= (-1)^n\vec{W}+\vec{w}$,
where $\vec{w}=(\vec{u}_{2}-\vec{u}_{3})/2$ and 
$\vec{W}=(\vec{u}_{2}+\vec{u}_{3})/2$,
the spin-orbit term can then be written as
\begin{equation}
\label{eqnso}
H'=\sum_{n}i ( c_{n,1}^{\dagger }\vec{w}_{n}\cdot \vec{
\sigma}c^{}_{n+1,1} 
+c_{n,2}^{\dagger }\vec{w}_{n+1}\cdot \vec{\sigma} c^{}_{n+1,2}) + {\rm H.c.},
\end{equation}
consisting of a uniform and 
an alternating contribution,
\[
c_{n,1/2}^{\dagger }\vec{w}_{n}\cdot \vec{\sigma}c^{}_{n+1,1/2}
=c_{n,1/2}^{\dagger }\vec{w}\cdot \vec{\sigma}c^{}_{n+1,1/2}\pm (-1)^{n}
c_{n,1/2}^{\dagger }\vec{W}\cdot \vec{\sigma}c_{n+1,1/2} .
\]
The resulting model is studied in the remainder of this section
for the special case that both $\vec w$ and $\vec W$ are parallel to 
 the uniform magnetic field.
Note that $\vec w$ corresponds to the leading SO vector $\vec \lambda$
 in the low-energy theory, while $\vec W$ is related to the subleading vector $\vec
\lambda'$, see equation (\ref{las}). Moreover, away from half-filling, following
standard reasoning, one may expect that the alternating terms average out.
Nevertheless, they will be kept below, but we indeed confirm that in large 
systems they lead only to small effects.
When the above gauge transformation is applied again, we arrive at a
two-leg Hubbard ladder (\ref{eqnhubb}), but
with the on-chain hoppings $t$ carrying both a uniform phase $\exp(i\phi_u)$
and an alternating phase $\exp(i\phi_a)$ determined by $\vec w$ and $\vec W$, 
respectively, see figure \ref{ladderfas}.

\begin{figure}
\begin{center}
\epsfysize=12cm
\epsfbox{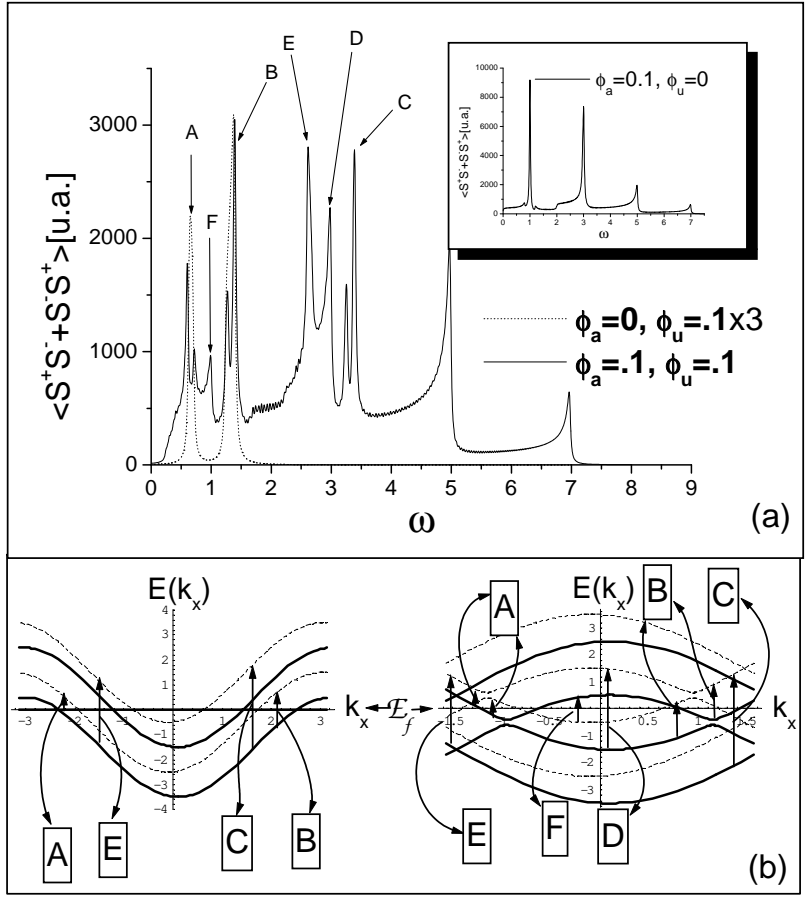}
\end{center}
\caption{ \label{esraltu}
a) ESR spectrum and b) dispersion relation of the non-interacting model for
$t= t_\perp=1$ and different 
phase configurations. The dotted (full) lines in (b) correspond to spins down (up).}
\end{figure}

Let us first discuss the noninteracting case.
For $U=0$, the solution of equations (\ref{eqnhubb}) and 
(\ref{eqnso}) is straightforward.
The eigenenergies are given by ($a$ is the lattice constant)
\begin{eqnarray*}
 E_{1\pm }(k_{x})&=&
-\sqrt{( 2t\cos (k_{x}a+\sigma \phi _{u})) ^{2}\pm
4tt_{\perp }\cos (k_{x}a+\sigma \phi _{u})\cos (\phi _{a})+t_{\perp }^{2}} \\
E_{2\pm }(k_{x})&=&
\sqrt{( 2t\cos (k_{x}a+\phi _{u})) ^{2}\pm 
4tt_{\perp }\cos (k_{x}a+\phi _{u})\cos (\phi _{a})+t_{\perp }^{2}} 
\end{eqnarray*}
where $\sigma=\pm 1$ for up or down electrons respectively.
In figure \ref{esraltu} we show the non-interacting $T=0$ ESR spectrum 
in the presence of  both the uniform and  the alternating phase. 
The main excitations contributing to each peak are
sketched in figure \ref{esraltu}(b). The uniform phase case (see dotted line) 
produces a splitting of the principal Zeeman peak. Due to the different 
parity of the $k=0$ and $k=\pi$ bands, for the pure uniform 
case there is no ESR interband transition. When the alternating 
phase is turned on, the folding of the Brillouin zone produces 
peaks at higher energies. As the parity symmetry is 
broken, interband transitions are now possible. The opening of the 
small gap due to the 
alternating phase is manifested in the splitting 
seen in peaks A, B and C. The broad structure (F and D) between 
the highest peaks correspond to transitions between branches with opposite 
curvature.

\begin{figure}
\begin{center}
\epsfysize=10cm
\epsfbox{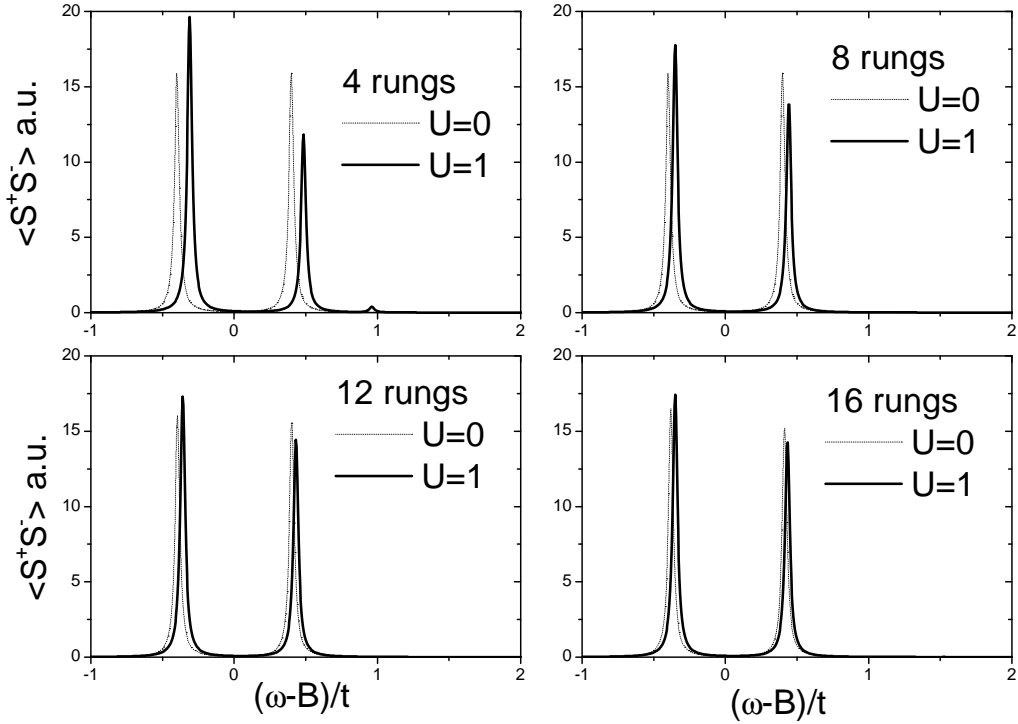}
\end{center}
\caption{ESR spectra for periodic quarter-filled ladders with a 
uniform phase $\phi_{u}=0.1$, $t=1$ and $t_{\perp}=1.2$.}
\label{perunifdmrg}
\end{figure}

Next we discuss results for $U>0$ using the density matrix 
renormalization group (DMRG) technique \cite{whiteprl,peschel}.
We have used periodic boundary conditions keeping 256 
states and the finite-system algorithm. 
The ESR spectrum (\ref{int1}) at $T=0$ has been calculated 
using the dynamical DMRG technique \cite{hallberg,phil}
for a two-leg ladder with uniform phase $\phi_u$ only.
Additional small alternating phases $\phi_a$
did not change the spectrum significantly. 
In figure \ref{perunifdmrg} we present these results for a quarter filled 
chain and $t_{\perp }=1.2$ and $t=1$. These parameters are used in order to 
clearly see the effect of the SO  coupling, since for this filling the 
excitations occur  near the linear 
term of the cosine bands for system sizes that are multiple of 4. 
The effect of the correlations mainly consists 
of a small shift of the peaks, but the double peak spectrum is
preserved.  Therefore the numerical results lend support to the
basic prediction of the analytical low-energy theory, and show
that the expected double-peak ESR spectrum is stable with respect
to the above-mentioned perturbations.

\section{Conclusions}
\label{sec6}

We have reviewed the analysis of the ESR spectrum produced by the spin-orbit 
coupling in SWNTs. 
The effective field theory analysis shows that at low energy the SO interaction only
acts in the spin sector and the single Zeeman peak, characteristic of a system with
$SU(2)$ spin symmetry, splits into two peaks with no broadening. 
This result relies in an essential way on the property of spin-charge 
separation characteristic of the Luttinger liquid state realized in the SWNT. 
Thus the observation of such a splitting would point 
to this elusive feature.
 
The field theory analysis has been complemented by the exact 
numerical calculation of 
the ESR spectrum for Hubbard models on a single chain and on a two-leg ladder.
While these models are not able to describe reliably the charge transport properties 
of SWNTs due to neglect of the long-range tails of the Coulomb interaction, they are 
appropriate descriptions of this spin problem.
The numerical analysis confirms the field theory predictions, including
the validity of the approximations involved in their derivation.
In addition, it reveals additional structure in the spectrum due to 
band curvature and higher energy processes, which are not captured 
by the field theory approach.

\ack
We thank C.A. Balseiro for the collaboration in the first stage of this work and
L. Forr{\'o} for valuable discussions. Support by the DFG under the 
Gerhard-Hess program, and by the project PICT 99 3-6343 is acknowledged.

\end{document}